\documentclass[aps,pra,amsfonts,amssymb,amsmath,twocolumn]{revtex4}
\usepackage{graphicx,hyperref}
\def\be{\begin{equation}}
\def\ee{\end{equation}}
\def\ba{\begin{eqnarray}}
\def\ea{\end{eqnarray}}
\def\sgn{{\rm sgn}}
\begin{document}

\title{Ground states of hard-core bosons in one dimensional periodic potentials}

\author{Yuan Lin}
\affiliation{Institute of Physics, Chinese Academy of Sciences,
Beijing 100080, China}
\author{Biao Wu}
\affiliation{Institute of Physics, Chinese Academy of Sciences,
Beijing 100080, China}

\begin{abstract}
With Girardeau's Fermi-Bose mapping, we find the exact ground states
of hard-core bosons residing in a one dimensional periodic
potential. The analysis of these ground states shows that when the
number of bosons $N$ is commensurate with the number of wells $M$ in
the periodic potential, the boson system is a Mott insulator whose
energy gap, however, is given by the single-particle band gap of
the periodic potential; when $N$ is not commensurate with $M$, the system
is a metal (not a superfluid). In fact, we argue that there may be
no superfluid phase for any one-dimensional boson system in terms
of Landau's criterion of superfluidity. The Kronig-Penney potential
is used to illustrate our results.
\end{abstract}

\maketitle
\section{INTRODUCTION}
A Tonks-Girardeau (TG) gas is a one dimensional strongly correlated
system consisting of bosons with hard-core interaction. This model of
a one dimensional gas was first studied by Tonks\cite{Tonks36},
who treated it as a classical gas. Later studies were
nevertheless focused mostly on its properties as a quantum boson
gas\cite{Bijl,Nagamjya}. As the milestone development in this model,
Girardeau\cite{Gir60,Gir65} found that this boson gas can be mapped
into a spinless fermion system and obtained all of its  eigenstates.

This idealized TG gas had only been a curiosity of theorists until
the realization of Bose-Einstein condensation with dilute atomic
gases in 1995\cite{Morsch06,Andersen-RMP04,Bongs-RPP}. It was soon
noticed that the TG gas could be realized experimentally with a
Bose-Einstein condensate trapped in tightly-confined
waveguides\cite{Olshanii98}. This work renewed interests in the TG
gas and led to a series of further theoretical
studies\cite{Petrov00,Wright,Gir00,Lapeyre02,Das,Forrester,Minguzzi,Campo}.
In particular, Lapeyre {\it et al.} studied the momentum distribution of a
harmonically trapped TG gas\cite{Lapeyre02} and
Forrester\cite{Forrester} {\it et al.} obtained highly accurate results
on the TG gas's long-range off-diagonal behavior\cite{Lenard64,Lenard66}.
Finally in 2004, the TG gas was first realized with a Bose-Einstein
condensate trapped in two perpendicular optical
lattices\cite{Paredes-nature,Kinoshita}. An overview of studies in
this interesting subject can be found in a recent review by Yukalov
and Girardeau\cite{Yukalov05}.

In this article we study the TG gas in a periodic potential. We
focus on the case that the hard-core interaction between bosons is a
repulsive infinite contact interaction. By applying Girardeau's
Fermi-Bose mapping\cite{Gir60,Gir65}, this Boson gas can be mapped
into a spinless free Fermi system. As a result, we can obtain all
the eigenstates and eigen-energies of this TG gas. We are
particularly interested in its ground properties. Our analysis of
the ground states shows that such a TG gas has two quantum phases, Mott
insulator and metal. When the number of bosons $N$ in the gas is
commensurate with the number of wells $M$ in the periodic potential,
the boson system is a Mott insulator; however, its energy gap is
given by the single-particle band gap of the periodic potential.
When $N$ is not commensurate with $M$, the system is a metal.
We emphasize that this boson metal is not a superfluid as it
can not support any superflow with finite critical velocity.
In fact, in this sense there may be no superfluid phase for
any one-dimensional boson systems including soft-core bosons.

The Kronig-Penney potential is used to illustrate our results.
Various properties of the ground state are computed, such as pair
distribution function, single-particle density matrix, and momentum
distribution. Before we proceed with full discussion, we note
that the hard-core bosons in a lattice have been actively studied
under single-band approximation with the Jordan-Wigner
transformation\cite{Rigol1,Rigol2,Gangardt,Rigol3}. It
is also studied as a limiting case in Ref.\cite{buchler}

\section{TG GAS IN PERIODIC POTENTIAL}
Normally, the boson in a TG gas has an
``impenetrable'' hard core characterized by a radius of $a$. In this
study, we focus on the case that the hard core is a point with no
radius, that is, the inter-particle interaction is given by
\be
U(x)=\left\{
\begin{array}{ll}
\infty &~~~~x=0\,,\cr 0&~~~~x\neq 0\,.
\end{array}
\right.
\ee
Such an interaction is equivalent to a constraint on the
wave function $\psi (x_1,\cdots,x_N)$.
\begin{eqnarray}
\psi=0~~~~~{\rm if}~~~~ x_i = x_j,~~ 1\leqslant i<j\leqslant N.
\label{eq:constraint}
\end{eqnarray}
This means that this TG gas can be viewed as a group of ``free'' bosons
governed by the following Hamiltonian
\begin{equation}
\label{eq:ham}
\hat{H}=\sum_{j=1}^{N}\Big\{-\frac{\hbar^2}{2m}\frac{\partial^2}{\partial
x_j^2}+V(x_j)\Big\}\,,
\end{equation}
while its wavefunction is subject to the constraint in
Eq.(\ref{eq:constraint}). Based on the observation that the
constraint in Eq.(\ref{eq:constraint}) is automatically satisfied by
any wavefunction of a Fermi system due to its antisymmetry,
Girardeau\cite{Gir60,Gir65} found a natural mapping that allows one
to construct wavefunctions for a TG gas from wavefunctions of a free
Fermi system. The mapping is
\be \label{eq:bf}
\psi^B(x_1,x_2,\cdots,x_N)=A\psi^F(x_1,x_2,\cdots,x_N)\,
\ee where
$\psi^B$ is the Bose wavefunction and $\psi^F$  the Fermi
wavefunction. $A$ is called unit antisymmetry function which is
defined as \be A(x_1,x_2,\cdots,x_N)=\prod_{i>j}\sgn(x_i-x_j)\,, \ee
where $\sgn$ is the sign function. This Fermi-Bose mapping is
one-to-one and, therefore, the energy spectrum of a TG gas of
infinite contact potential is the same as a free spinless Fermi
system\cite{Gir60,Gir65}.

In this article, we study the case where the external potential
$V(x)$ is periodic, $V(x)=V(x+d)$, with $d$ being the period. Due to
the Fermi-Bose mapping (\ref{eq:bf}), we consider first the system
of $N$ free spinless fermions residing in this periodic potential.
According to the basic knowledge of solid state
physics\cite{Mermin}, when the number of fermions $N$ in the gas is
commensurate with the number of wells $M$ in the periodic potential,
the fermions can fill up exactly $N/M$ Bloch bands and thus the
Fermi system is an insulator with its energy gap determined by the
periodic potential. When $N$ is not commensurate with $M$, the
fermions can fill up the bands only partially and the Fermi system
is a metal.

Because of the Fermi-Bose mapping (\ref{eq:bf}), the TG gas has the
same energy spectrum as this free Fermi system. This allows us to
conclude immediately that when $N$ is commensurate with $M$, the TG
gas is an insulator. However, we emphasize that this insulator is a
Mott insulator unlike its mapping target, the free Fermi system,
where the insulator is a band insulator. This leads to a quite
peculiar situation, the energy gap of a Mott insulator is completely
dictated by the periodic potential and is given by the single-particle
band gap. For the other situation where
$N$ is not commensurate with $M$, the TG gas is a metal.

\begin{figure}[!htb]
  \includegraphics[width=7cm]{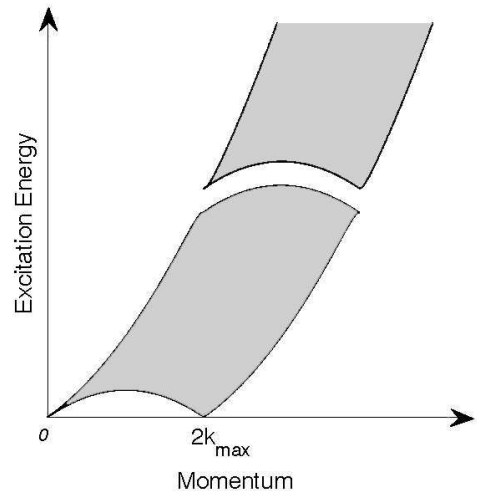}
  \caption{Excitation spectrum of a TG gas in a periodic potential
for the non-commensurate case. The shade makes out all the possible
excitations. $k_{\rm max}$ is the largest
possible value of $k$, similar to the Fermi wavevector in the fermion
case.}
\label{excitation}
\end{figure}

We emphasize that this Bose metal is a real metal not a superfluid
according to Landau's criterion of superfluidity\cite{Landau}.
Landau's criterion is that the system is a superfluid if it has
phonons as its only low energy excitations.  Plotted in Fig.\ref{excitation}
is the excitation spectrum of a TG gas in a periodic potential
for the non-commensurate case, which is marked out by the shade.
It is clear from this figure that there are two kinds of
low energy excitations, one at $k=0$ and the other at $k=k_{\rm max}$.
The low energy excitation at $k=0$ may be barely called phonon
since the phonon relation $\epsilon=ck$ exists only at the limit
$k\rightarrow 0$ and the excitation spectrum immediately spreads out
at $k\neq 0$. The most destructive is the low energy excitations
at $k_{\rm max}$, which are not phonons. The presence of these
low excitations make it vanishing the minimum slope of $\epsilon(k)/k$,
which gives the critical velocity of possible superfluidity.
Therefore, we can conclude that this TG gas can not support
a superflow with finite critical velocity and it is not a superfluid.

What is more interesting is that the TG gas is not an isolated case.
According to Ref.\cite{Lieb}, a boson gas with finite repulsive
$\delta$-function interaction has a very similar excitation
spectrum as our TG gas. This means that such a gas is no
superfluid, either. This prompts us
to speculate any one-dimensional boson system is not a superfluid.
We are aware that there are different definitions of
superfluidity\cite{Lieb2005Book}. In particular, according to
Leggett\cite{Leggett}, our TG gas (in his version it is spinless
fermion) is a superfluid since
it has the so-called non-classical rotational inertia (NCRI).
Because of this, we state again that we are talking
about superfluid in the sense of Landau's criterion.

We now construct explicitly the exact eigenfunctions of this TG gas.
The single-particle eigenstate $\varphi_{n,k}(x)$ is a Bloch
state that satisfies the Schr\"odinger equation
\begin{eqnarray}
[-\frac{\hbar^2}{2m}\frac{\partial^2}{\partial x^2}+
V(x)]\varphi_{n,k}(x)=E_{n,k}\varphi_{n,k}(x)\,.
\end{eqnarray}
The notation $n$ is for band index and $k$ is the Bloch wave vector
in the first Brillouin zone. For convenience, we
use $\alpha=\{n,k\}$ to denote both the band index and the Bloch wave vector.
Then according to Eq.(\ref{eq:bf}), the Bose eigenfunction is given by
the Slater determinant
\begin{eqnarray}
&&\psi^B(x_1,x_2,\cdots,x_N)=\nonumber\\
&=&\frac{A}{\sqrt{N!}}\left|
\begin{array}{cccc}
\varphi_{\alpha_1}(x_1)&\varphi_{\alpha_2}(x_1)&\cdots&\varphi_{\alpha_N}(x_1)\\
\varphi_{\alpha_1}(x_2)&\varphi_{\alpha_2}(x_2)&\cdots&\varphi_{\alpha_N}(x_2)\\
\vdots & \vdots & \vdots & \vdots\\
\varphi_{\alpha_1}(x_N)&\varphi_{\alpha_2}(x_N)&\cdots&\varphi_{\alpha_N}(x_N)
\end{array}
\right|\,.
\end{eqnarray}
As one can check, this symmetric wavefunction satisfies both the Schr\"odinger
equation with the Hamiltonian
(\ref{eq:ham}) and the constraint (\ref{eq:constraint}).
The corresponding eigen-energy is
\begin{eqnarray}
E=\sum_{j=1}^{N}E_{\alpha_j}\,.
\end{eqnarray}
For the ground state, the $N$ eigenfunctions in the Slater
determinant are for the $N$ lowest eigenstates. For convenience, we
shall assume that both $N$ and $M$ are odd in the following
discussion.

\section{TG GAS IN The Kronig-Penney POTENTIAL}
\begin{figure}[!htb]
  \includegraphics[width=7cm]{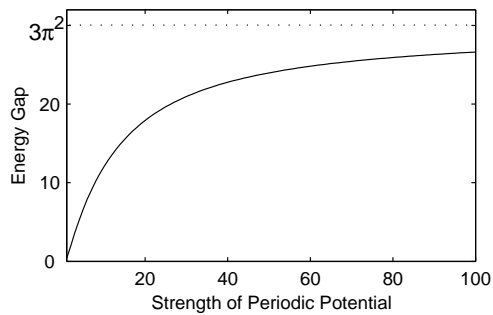}
  \caption{Width of the energy gap between the first and the second Bloch band
  as a function of the strength $V$ of periodic potential. This is also the
energy gap of the Mott insulator when the number of bosons in the TG gas
is equal to the number of wells in the periodic potential.}
\label{Energy gap}
\end{figure}
We now use the Kronig-Penney (KP) potential to illustrate our
general results of the TG gas in the last section.
The Kronig-Penney potential is given by
\begin{eqnarray}
V(x)=\gamma\sum_j\delta(x-jd)\,
\end{eqnarray}
where $\gamma$ is the strength of the $\delta$-potential. We follow
the standard procedure in solid state physics \cite{Mermin} by
placing the system in a box whose size is $L=Md$ and imposing the usual
periodic boundary condition. If we
use $d$, the period of the potential, as the unit of distance and
$2md^2/\hbar^2$ as the unit of energy, the Schr\"odinger
equation can be written as
\begin{eqnarray}
[-\frac{\partial^2}{\partial
x^2}+V\sum_{j=0}^{M-1}\delta(x-j)]\varphi_\alpha (x)=E_\alpha
\varphi_\alpha(x)
\end{eqnarray}
where $V=2m\gamma d/\hbar^2$. The eigenfunction $\varphi_\alpha$ is
given by
\begin{eqnarray}
\varphi_\alpha(x)= \label{eq:1}
  \left\{
    \begin{aligned}
        &f(x)=A_\alpha e^{ip_\alpha x}+B_\alpha  e^{-ip_\alpha x}~~~x\in[0,1)\\
        &e^{iks}f(x-s)~~~x\in[s,s+1)\\
        &~~~~~~~~~~~~~~~~~~~~~~~(s=1,2,\cdots,M-1)
     \end{aligned}
  \right.
\end{eqnarray}
with the coefficients determined by
\begin{eqnarray}
\label{eq:2}
  \left\{
    \begin{aligned}
         &k=\frac{2\pi l}{M}~~~(l\in 0,\pm1,\cdots, \pm \frac{M-1}{2})\\
         &\cos p_\alpha +\frac{V}{2p_\alpha }\sin p_\alpha =\cos k\\
         &A_\alpha=\frac{\sqrt{2}e^{-ip_\alpha /2}p_\alpha \sin [(p_\alpha+k )/2]}
{\sqrt{[(2p_\alpha ^2+V)\sin p_\alpha -p_\alpha V\cos p_\alpha ]M\sin p_\alpha }}\\
         &B_\alpha=\frac{\sqrt{2}e^{ip_\alpha /2}p_\alpha\sin [(p_\alpha-k )/2]}{\sqrt{[(2p_\alpha ^2+V)\sin p_\alpha
-p_\alpha V\cos p_\alpha ]M\sin p_\alpha }}.
     \end{aligned}
  \right.
\end{eqnarray}
The corresponding eigen-energy is $E_\alpha=p_\alpha^2$.

As we have demonstrated in the last section, when a TG gas of
exactly M bosons is placed in the KP potential, the system is a Mott
insulator. The energy gap between the first and the second Bloch
band is plotted in Fig.1. This figure shows that this gap initially
increases with the lattice strength $V$ and eventually saturates at
the energy difference between the first excited state and the ground
state in a square well potential.

\subsection{PAIR DISTRIBUTION FUNCTION}
The pair distribution function, normalized to $N(N-1)$, is
defined as
\begin{eqnarray}
&&D(x_1,x_2)=N(N-1)\int|\psi_0^B(x_1,\cdots,x_N)|^2 dx_3\cdots
dx_N\nonumber\\
&&~~~=\frac{1}{2}\sum_{\alpha,\alpha'=0}^{N-1}
|\varphi_{\alpha}(x_1)\varphi_{\alpha'}(x_2)-\varphi_{\alpha}(x_2)\varphi_{\alpha'}(x_1)|^2.
\label{eq:pair}
\end{eqnarray}
It is the joint probability of finding simultaneously one
atom at $x_1$ and another atom at $x_2$. When $x_1=x_2$, the
pair distribution function vanishes by antisymmetry, reflecting
physically the impenetrable hard-core interaction between boson
particles.
\begin{figure}[!htb]
  \includegraphics[width=8cm]{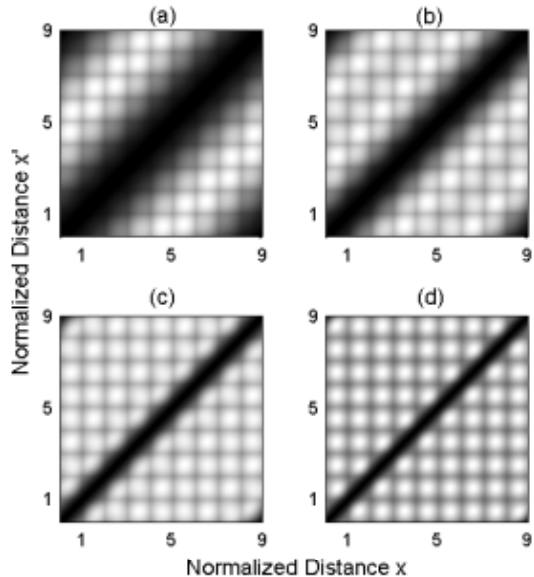}
  \caption{Gray-scale plots of the  pair
distribution function $D(x_1,x_2)$. $V$=1. (a) $N$=2, $M$=9;
(b) $N$=3, $M$=9; (c) $N$=6, $M$=9; (d) $N$=9,
$M$=9.}
\label{PairDis93}
\end{figure}

Figure \ref{PairDis93} shows gray-scale plots of the  pair
distribution function $D(x_1,x_2)$ for different numbers of
particles of $M$=9. Some qualitative features of the pair distribution
function are observed. First, the figure shows that
$D(x_1,x_2)$ vanishes at contact $x_1=x_2$, reflecting
the impenetrability between the particles. Second, the periodicity due to
the Kronig-Penney potential is apparent.
Third, with the increase of the particle number,
the black diagonal stripe becomes thinner. This indicates
that the averaged distance between particles decreases.

\subsection{REDUCED SINGLE-PARTICLE DENSITY MATRIX}
\begin{figure}[!htb]
  \includegraphics[width=8cm]{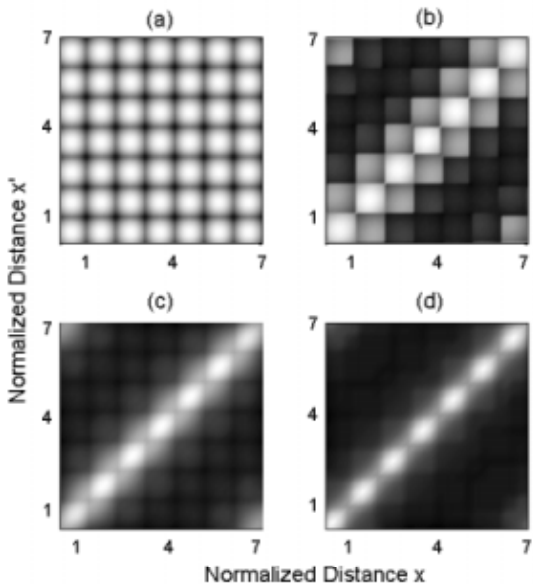}
  \caption{Gray-scale plots of the reduced density matrix
$\rho(x,x')$ of the TG gas. $V$=1. (a) $N$=1, $M$=7; (b) $N$=3,
$M$=7; (c) $N$=5, $M$=7;(d) $N$=7, $M$=7.}
\label{ReducedSPD}
\end{figure}
The reduced single-particle density matrix is given by
\begin{eqnarray}
\rho(x,x')=&&N\int\psi_0^B(x,x_2,\cdots,x_N)\nonumber\\
&&\psi_0^B(x',x_2,\cdots,x_N)dx_2\cdots dx_N.
\label{eq:reduced}
\end{eqnarray}
The existence of the off-diagonal long-range order in this matrix
indicates the onset of Bose-Einstein condensation \cite{Penrose,
Yang}. Its diagonal term $\rho(x)=\rho(x,x'=x)$ is the
single-particle density and is normalized to $N$,
\begin{eqnarray}
\int \rho(x,x)dx=N\,.
\end{eqnarray}

The multidimensional integral in Eq.(\ref{eq:reduced}) is evaluated
numerically by Monte Carlo integration. The results are shown in
Fig.\ref{ReducedSPD}. The relative darker area of the
diagonal in the figure is due to $\delta$-potentials, which tend to repulse the
particles away. The particles like to stay between
$\delta$-potentials, i.e., the brighter diagonal area.

\begin{figure}[!htb]
  \includegraphics[width=8cm]{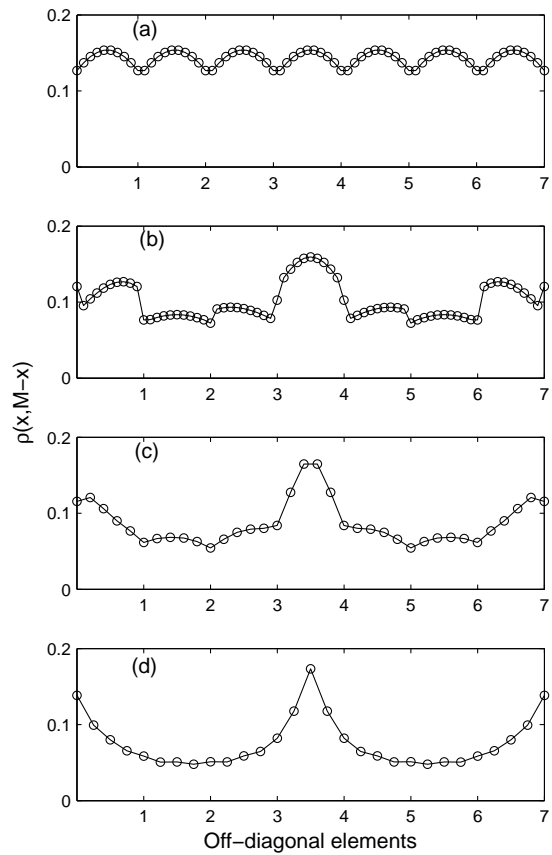}
  \caption{Off-diagonal elements $\rho(x,M-x)$ of
the reduced density matrix. $V=1$. (a) $N$=1, $M$=7; (b)
$N$=3, $M$=7; (c) $N$=5, $M$=7; (d) $N$=7, $M$=7.} \label{Dig}
\end{figure}

The gray scale in Fig.\ref{ReducedSPD} indicates that
 the off-diagonal elements decrease as $N$ increases.
To see this more clearly, we have plotted the off-diagonal
elements $\rho(x,M-x)$ of the reduced density matrix in
Fig. \ref{Dig}.  It is clear
from the figure that as $N$ increases and the interaction in the
system gets stronger, the off-diagonal elements decrease relative to
the diagonal. This indicates that the interaction suppresses the
off-diagonal terms. It has been proved that there is no real
Bose-Einstein condensate in the uniform Tonks-Girardeau gas
\cite{Lenard64,Lenard66,Forrester}.

\subsection{MOMENTUM DISTRIBUTION}
\begin{figure}
  \includegraphics[width=8cm]{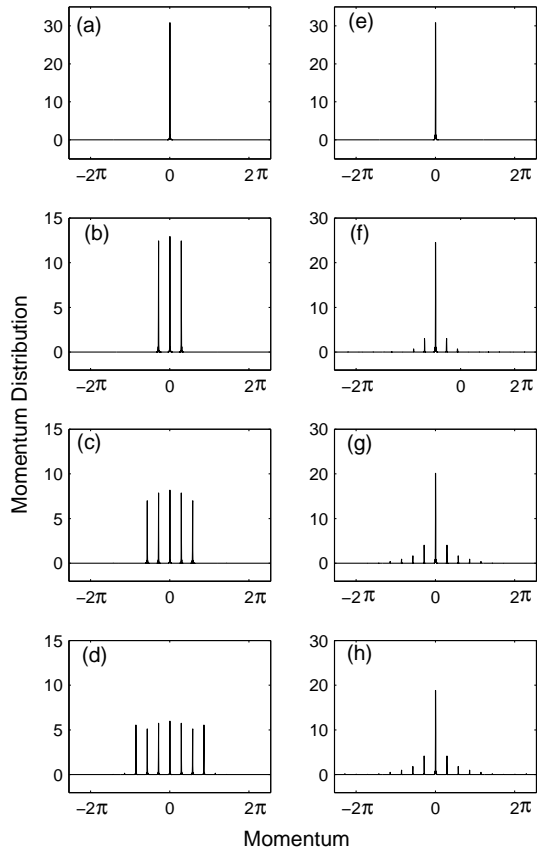}
  \caption{Normalized momentum distribution function $\varrho(k)$.
$V=1$. $M=7$. (a-d) Free Fermi
gas for $N$=1,3,5,7 (from top to bottom); (e-h)
TG gas for $N$=1,3,5,7 (from top to bottom).}
\label{momentum}
\end{figure}
The Bose-Einstein condensation of a Bose system is often
characterized by a macroscopic number of bosons occupying the zero
momentum state. It is therefore meaningful to look at the momentum
distribution of our TG gas in the ground state. The momentum
distribution $\varrho(k)$ is related to the reduced density matrix
and given by
\cite{Lapeyre02}
\begin{eqnarray}
\varrho(k)=\frac{1}{2\pi N}\int_{-\infty}^{+\infty} dx
\int_{-\infty}^{+\infty} dx' \rho(x,x')e^{-ik(x-x')},
\end{eqnarray}
which is normalized to one. As noted by Girardeau\cite{Gir60,Gir65},
the Fermi-Bose mapping in Eq.(\ref{eq:bf})
becomes simply $\psi^B=|\psi^F|$ for the ground state. This shows
that even though the density distribution $|\psi|^2$ is
the same for the TG gas and its mapping counterpart, free Fermi
gas, their momentum distributions are different. It is therefore
also interesting to compare the momentum distributions for
these two different systems.

In Fig.\ref{momentum}, we have plotted the momentum distributions
for both the TG gas and the free Fermi gas in a periodic potential.
The number of wells are  $M=7$ and the strength of the periodic
potential is $V=1$. The left column (a-d) of Fig.\ref{momentum}
is for the Fermi gas and the right column (e-h) is for the TG gas.
Since higher order Bragg peaks, e.g., the ones at $\pm 2\pi$,
are very small, the plots in Fig.\ref{momentum} focus on the central
Bragg peak, which consists of discrete small peaks due to
the finite size of the systems.

We first look at the right column of Fig.\ref{momentum}.
For $N=1$, which can be regarded as the free boson case,
we see that the Bragg peak is completely located at $k=0$.
As the number of bosons increases, the interaction
in the system gets stronger. This causes the Bragg peak
to spread out as seen in the right column of Fig.\ref{momentum}.
When $N=M$,  the peak has a width of almost a whole Brillouin zone.
It is interesting to compare the TG gas with the free Fermi gas.
As shown in the left column of Fig.\ref{momentum},
the Bragg peaks are much ``fatter'' than the TG gas (except $N=1$).
This demonstrates that even a free Fermi gas has a broader
momentum distribution than a strongest interacting boson gas.

\section{SUMMARY AND CONCLUSIONS}
In summary, we have studied the ground-state properties of a
Tonks-Girardeau gas in a periodic potential with the Girardeau's
Fermi-Bose mapping. We found that such a TG gas is a metal when the
number of particle $N$ is not commensurate with the number of wells
$M$ in the periodic potential; it is a Mott insulator when $N$ is
commensurate with $M$. What is more interesting is the energy gap in
the Mott insulator is given by the single-particle band gap of the
periodic potential.  To illustrate our results, we have employed the
Kronig-Penney potential to compute various properties of the ground
state, such as pair distribution function, single-particle density
matrix, and momentum distribution.

\section{Acknowledgement}
We are supported by the ``BaiRen'' program of the Chinese Academy of
Sciences, NSF of China (10504040), and the 973 project of China
(2005CB724500).

\end{document}